# Excitation of relaxation oscillations in a semiconductor superlattice by incident waves: efficient terahertz harmonics generation

**Anatoly A Ignatov**

Institute for Physics of Microstructures, Russian Academy of Sciences, GSP-105, Nizhny Novgorod, 603950 Russia

E-mail: ign@ipm.sci-nnov.ru

**Abstract.**
Generation of terahertz harmonics by frequency multiplication with a semiconductor superlattice due to an excitation of relaxation oscillations by incident waves is investigated theoretically. It is shown that the relaxation oscillations excitation becomes feasible if the superlattice dc resistance is low enough in comparison with a characteristic radiation impedance of the external waveguide system. The power of the generated harmonics as a function of the incident wave power demonstrates a threshold-like behavior at a specific input power level dependent on the superlattice peak current. We demonstrate that for typical superlattice parameters the roll-of frequency of the generated harmonics is mostly specified by the plasma frequency of electrons in a superlattice miniband. We argue that an increase of the superlattice miniband widths could essentially enhance both the efficiency and the spectral range of the generated terahertz harmonics.

# 1. Introduction.

Investigation of semiconductor superlattices proposed by Esaki and Tsu in their seminal paper [1] has attracted a great deal of interest from many experimental and theoretical research groups. Among numerous topics, investigation of Bloch oscillations of electrons [2], infrared spectroscopy of electron states [3], study of the nonlinear transport in high ac/dc electric fields [4], works on the nonlinear spatiotemporal dynamics of carriers in superlattices [5] and observation of the Bloch gain in quantum cascade lasers [6,7] can be especially emphasized.

On the other hand, research into terahertz technology and devices is now becoming increasingly important [8]. In this respect several novel models of the superlattice ac electron response such as parametric amplification of the terahertz signals [9], Bloch gain in dc-ac-driven superlattices in the absence of electric domains [10], and terahertz Bloch oscillator with a modulated bias [11] have been suggested.

Meanwhile, a semiclassical theory of the high-frequency response of superlattices suggests that these structures may be of considerable interest for the frequency multiplication of intense electromagnetic signals in the terahertz frequency band [12-14]. First experiments initiated by this theory clearly demonstrated that superlattice devices were quite suitable for generation of sub-terahertz frequency waves by frequency multiplication of the microwave radiation [15]. Moreover, frequency doubling and tripling of terahertz radiation from the free-electron laser with a GaAs/AlAs superlattice was shown experimentally [16].

Later, frequency triplers for generation of sub-terahertz waves based on a novel semiconductor superlattice quasi-planar design have been exploited in papers [17, 18]. Experiments carried out in these papers have revealed a curious threshold-like dependence of the generated third harmonics power as a function of the pump microwave power.

Recently, frequency multipliers based on quasi-planar superlattice devices as nonlinear elements have been developed as radiation sources for terahertz laboratory spectrometers [19-21] and for phase locking of quantum cascade lasers [22]. The multipliers of this type having the input wave frequencies of about 100-250 GHz were able to generate up to 11th [19] or even up to 35th [21] harmonics with sufficient for terahertz spectroscopy and frequency locking output powers.

In this article, we treat generation of the terahertz harmonics by frequency multiplication in semiconductor superlattices irradiated by incident waves. A non-resonant equivalent circuit taking into account a superlattice capacitance, a series resistance, and a current of electrons capable to perform Bloch oscillations [23, 24] has been employed to define self-consistently the conversion efficiency and the power spectrum of the generated harmonics. Making use a system of nonlinear equations derived on the bases of this model we investigate the high-frequency dynamics of electron response in a superlattice both analytically and numerically.

We show that generation of ultra-short electromagnetic pulses in the superlattice can occur due to specific relaxation oscillations excited by incident low-frequency monochromatic waves. This gives rise to an ultra-broad spectrum of the emitted terahertz harmonics. Relaxation oscillations we study do not require any positive feedback provided by external resonators such as that ones applied to resonant tunneling diodes [25].

We have found that the dependence of the generated harmonics powers as a function of the incident wave power shows a threshold-like behavior at specific values of the input power. Our study demonstrates that the roll-of frequency of the generated harmonics is mostly specified by the plasma frequency of electrons in a superlattice miniband. We argue that by judicious increase of the superlattice miniband width one may be able to enhance both the efficiency and the spectral range of the generated terahertz harmonics.

## 2. Electro-dynamical model

For the qualitative description of the interaction of incident electromagnetic waves with our superlattice device we consider an example of the waveguide system depicted in figure 1(a). The waveguide system of this kind has been used in Schottky diode terahertz harmonics generators [26]. It consists of the input waveguide that is smoothly narrowed to an uprising central ridge. The superlattice device is placed at the end of the ridged part of the waveguide. To the right of the ridge, a horn antenna is placed serving as a high-frequency harmonics emitter.

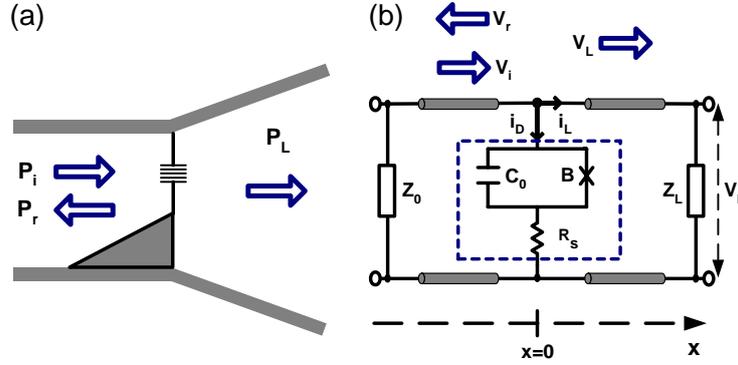

**Figure 1.** (a) Microwave radiation coupled to a vertical semiconductor superlattice device by a ridged waveguide; $P_i$, $P_r$, and $P_L$ are the incident, reflected and emitted powers, respectively. (b) Equivalent circuit for a superlattice-based terahertz harmonics generator: $B$-miniband electrons capable to perform Bloch oscillations, $C_0$-superlattice capacitance, $R_S$-parasitic series resistance, $Z_0$ - impedance of the ridged waveguide, $Z_L$ - impedance of the horn antenna.

We apply the transmission line approach [27] for the description of interaction between the incident waves and the superlattice. The equivalent circuit for the superlattice coupled to the ridged waveguide and the horn antenna is presented in figure 1(b), where $Z_0$ is the characteristic impedance of the ridged waveguide, and $Z_L$ is the impedance of the horn antenna. We assume that the characteristic impedance of the ridged waveguide $Z_0$ and the horn antenna $Z_L$ are frequency independent.

According to the transmission line theory [27] electromagnetic waves in the region $x \leq 0$ can be described by the solution of the telegrapher's equations for simple waves

$$V = F_V\left(t - x/v_{ph}\right), \tag{1}$$

$$i = F_i\left(t - x/v_{ph}\right), \tag{2}$$

where $V$ and $i$ are the transmission line voltage and current waveforms, $F_{V,i}(x)$ are the arbitrary form single-valued functions, $v_{ph}$ is the phase velocity of the waves, $F_i = Y_0 sgn(v_{ph}) F_V$, $sgn(x)$ is the signum function, and $Y_0 = 1/Z_0$ is the transmission line conductance.

The boundary conditions for the incident $V_i, i_i$, reflected $V_r, i_r$, and emitted $V_L, i_L$ waves at $x = 0$ can be obtained from the Kirchhoff's circuit laws

$$V_L = V_i + V_r, \tag{3}$$

$$i_L + i_D(V_D) = i_i + i_r, \tag{4}$$

where $i_D(V_D)$ is the device current.

Taking into account that $i_i = Y_0 V_i$, $i_r = -Y_0 V_r$, $V_D = V_L$, and $i_L = Y_L V_L$ we find

$$V_i + V_r = V_D, \tag{5}$$

$$Y_0(V_i - V_r) = i_D(V_D) + Y_L V_D, \tag{6}$$

where $Y_L = Z_L^{-1}$ is the load conductance.

We note that in equation (6) the function $i_D(V_D)$ specifies nonlinear properties of any particular device connected to our transmission line. If the incident waveform $V_i(t)$ is known, equations (5) and (6) allow one to find the forms for both the reflected $V_r(t)$ and the transmitted $V_L$ waves. Consequently, equations (5) and (6) unable us to find self-consistently the conversion efficiency and the power spectrum of the generated harmonics by means of Fourier analyzes of their solution for any particular device.

In a special case of a small amplitude monochromatic wave incident on the device, for the power reflection coefficient $R$, the transmission coefficient $T$, and the absorption coefficients $A$ equations (5) and (6) yield:

$$R = \left|Y_0 - \left(\hat{Y}_D + Y_L\right)\right|^2 \left|Y_0 + \left(\hat{Y}_D + Y_L\right)\right|^{-2}, \tag{7}$$

$$T = 4Y_0 Y_L \left|Y_0 + \left(\hat{Y}_D + Y_L\right)\right|^{-2}, \tag{8}$$

$$A = 1 - R - T = 4Y_0 \operatorname{Re}(\hat{Y}_D) \left|Y_0 + \left(\hat{Y}_D + Y_L\right)\right|^{-2}, \tag{9}$$

where $\hat{Y}_D$ is the complex conductance of the device connected to the transmission line, $Y_0$ and $Y_L$ are the real values.

## 3. Constitutive relations

For description of the non-linear electron transport in a superlattice we use a semiclassical wave packet treatment of the electron motion in a superlattice miniband [1]. The energy spectrum of electrons in a miniband is taken in a tight-binding approximation. In this case the energy of electron motion along the superlattice axis $z$ can be written as

$$\varepsilon(p_z) = \frac{\Delta}{2}\left[1 - \cos\left(\frac{p_z d}{\hbar}\right)\right], \quad (10)$$

where $\Delta$ is the miniband width, $d$ is the superlattice period, $p_z$ is the quasimomentum of an electron along the superlattice axis.

The quasiclassical velocity of an electron moving along the superlattice axis $v_z(p_z)$ and the time derivative of the quasimomentum $dp_z/dt$ are given by the expressions

$$v_z(p_z) = \frac{\partial \varepsilon(p_z)}{\partial p_z} = v_0 \sin(p_z d/\hbar), \quad (11)$$

$$\frac{dp_z}{dt} = eE_z(t), \quad (12)$$

where $e$ is the electron elementary charge, $v_0 = \Delta d/2\hbar$ is the maximum velocity of electrons along the superlattice axis, and $E_z(t)$ is the time-dependent electric field directed along the superlattice axis.

On the basis of equations (10)-(12) for the mean electron velocity $v$ and the mean electron energy $\varepsilon$ two balance equations have been derived from the Boltzmann equation with a relaxation time approximation model of the collision integral [12, 28, 29]. For the normalized electron current through the superlattice $\bar{I}$, the normalized voltage across the superlattice $U$, and the normalized deviation of electron energy $\eta$ these equation can be written as

$$\tau \frac{d\bar{I}}{dt} = (1-\eta)U - \bar{I}, \quad (13)$$

$$\tau \frac{d\eta}{dt} = \bar{I}U - \eta, \quad (14)$$

where $\tau$ is the relaxation time for electron scattering, $i_B$ is the electron current through the superlattice, $\bar{I} = I/2$, $I = i_B/i_p$ is the normalized electron current, $j = i_B/A = env$ is the current density, $v$ is the mean velocity of electrons, $i_p = Aj_p$ is the peak electron current, $A$ is the superlattice cross-section aria, $j_p = env_p$ is the peak current density, $n$ is the density of electrons in the superlattice,

$$v_p = \frac{1}{2} v_0 \frac{I_1(\Delta/2kT)}{I_0(\Delta/2kT)} \quad (15)$$

is the electron peak velocity, $I_1(x)$ and $I_0(x)$ are the modified Bessel functions of first and zero order, $V_B$ is the voltage across the superlattice, $U = V_B/V_p$ is the normalized voltage, $E = V_B/L$ is the electric field strengths in the superlattice, $L$ is the superlattice length, $V_p = LE_p$ is the peak voltage, $E_p = \hbar\nu/ed$ is the peak electric field, $\nu = 1/\tau$ is the electron scattering frequency, $\eta = (\varepsilon - \varepsilon_T)/\varepsilon_p$ is the normalized deviation of the mean energy of electrons $\varepsilon$ from the thermal energy $\varepsilon_T = (\Delta/2)[1 - I_1(\Delta/2kT)/I_0(\Delta/2kT)]$, $\varepsilon_p = 2v_p\hbar/d$ is the characteristic energy of electrons in the superlattice miniband, and $T$ is the lattice temperature.

We note that equation (13) can be interpreted as the equation of motion for a particle with the energy-dependent effective mass in the presence of friction [29]. The first term in the right-hand part of equation (13) describes the linear dependence of the inverse electron effective mass as a function of the normalized energy $\eta$, while the second one is responsible for an exponential time decay of the current due to electron scattering. At the same time equation (14) for the normalized energy $\eta$ manifests the energy conservation law [29].

At small electric fields in the superlattice $U \ll 1$, $\eta \ll 1$ equations (13) and (14) result in the Drude formula for the complex ac miniband electron conductivity $\hat{\sigma}(\omega) = \sigma_0/(1 + i\omega\tau)$, where $\sigma_0 = en\mu_0$ is the static electron conductivity, and $\mu_0 = 2v_p/E_p$ is the static electron mobility. We note that the static conductivity $\sigma_0$ is proportional to the factor $I_1(\Delta/2kT)/I_0(\Delta/2kT)$ which is responsible for a "thermal saturation" of the miniband transport in a superlattice [30]. This factor side by side with equation for the maximum velocity of electrons $v_0$ gives rise to an essential increase of the static superlattice conductivity $\sigma_0$ with increasing of the miniband width $\Delta$ [30].

On the other hand, when the constant voltage $U_0$ (electric field $E_0$) is suddenly applied to the superlattice, equations (13) and (14) yield:

$$\tau^2 \frac{d^2 I}{dt^2} + \tau \frac{dI}{dt} + \left(1 + U_0^2\right) I = 2U_0. \quad (16)$$

Equation (16) describes an oscillatory transient evolution of the current $I$ (the mean velocity $v$) with the Bloch frequency $\Omega_B = eE_0 d/\hbar$ and the characteristic damping time $\tau_{damp} \approx \tau$ [28] to the steady-state Esaki-Tsu current-voltage (velocity-field) characteristics [1]

$$I = \frac{2U_0}{1 + U_0^2}. \quad (17)$$

Consequently, one can conclude that the excitation of the damped Bloch oscillations by the incident electromagnetic waves is inherently involved into the non-linear ac response of the superlattice described by equations (13) and (14).

## 4. Nonlinear ac response

Consider now that an intense wave $V_i(t)$ is incident on our superlattice device. This wave induces both electron and displacement currents in the superlattice device described by the equivalent circuit presented in figure 2(b).

According to this equivalent circuit the current through the superlattice $i_B$, the voltage drop across the superlattice $V_B$, the voltage drop across the load impedance $V_L$, and the total current through the superlattice device $i_D$ are govern by the Kirchhoff's equations

$$C_0 \frac{dV_B}{dt} + i_B = i_D, \qquad (18)$$

$$V_L = V_B + i_D R_S, \qquad (19)$$

where $C_0 = \varepsilon\varepsilon_0 A/L$ is the superlattice capacitance, $\varepsilon$ is the relative lattice dielectric constant, $\varepsilon_0$ is the electric field constant, and $R_S$ is the series resistance.

In this case equations of the electro-dynamical model (5) and (6) combined with the equivalent circuit equations (18) and (19) yield

$$\tau_d \frac{dU}{dt} = I_0(t) - I - YU, \qquad (20)$$

$$U_L = \overline{T}U_i + \overline{S}U, \qquad (21)$$

$$U_r = \overline{R}U_i + \overline{S}U, \qquad (22)$$

where $I_0(t) = (2Y_0 R_0 \overline{S})U_i(t)$ is the external driving current induced in the superlattice by the incident wave, $U_i(t) = V_i(t)/V_p$ is the normalized voltage of the incident wave,

$$\tau_d = \frac{\varepsilon\varepsilon_0}{\sigma_p} = \frac{\varepsilon\varepsilon_0 E_p}{j_p} \qquad (23)$$

is the dielectric relaxation time, $\sigma_p = j_p/E_p = \sigma_0/2$ is the characteristic static conductivity of the superlattice,

$$Y = (Y_0 + Y_L)R_0 \overline{S} \qquad (24)$$

is the normalized effective transmission line conductance, and $R_0 = V_p/i_p$ is the characteristic superlattice dc resistance.

The transmitted $U_L$ and reflected $U_r$ waves influenced by electrons in the superlattice are defined by equations (21) and (22), where

$$\bar{T} = 2Y_0 \left(Y_0 + Y_L + Y_S\right)^{-1}, \tag{25}$$

$$\bar{R} = \left(Y_0 - Y_L - Y_S\right)\left(Y_0 + Y_L + Y_S\right)^{-1}, \tag{26}$$

$$\bar{S} = Y_S \left(Y_0 + Y_L + Y_S\right)^{-1}, \tag{27}$$

and $Y_S = R_S^{-1}$ is the series conductance.

We note that the last right-hand term in equation (20) has been heuristically introduced in [31, 32] as a phenomenological shunting conductance of the superlattice related to an edge conductivity. Meanwhile, in our case this term describes the interaction of the superlattice with the external circuit playing a role of the normalized effective radiation conductance. It should be distinguished from the series conductance $Y_S$ giving rise to additional energy dissipation in the superlattice device.

In fact, the normalized radiation conductance $Y$ is presented by the contribution from both the radiation transmission line losses described by the factors $Y_0$ and $Y_L$ and the dissipative losses described by the factor $\bar{S}$. If our device is disconnected from the transmission line, i.e. if $Y_S \to 0$, $\bar{S} \to 0$, one has $Y \to 0$. In this case no radiation occurs from the electron current flow in the device. In the opposite case $Y_S \to \infty$, according to equations (20)-(27) one can neglect the influence of the series resistance on the superlattice ac response if $R_S \ll R_0 R_L / (R_0 + R_L)$.

The physical significance of equations (21) and (22) can be demonstrated by two opposite limiting cases. Firstly, for $R_S \to 0$, $Y_S \to \infty$ (no voltage drop across the series resistance $R_S$) $\bar{S} \to 1$, $\bar{T} \to 0$ and $\bar{R} \to -1$. In this case $U_L = U$ and $U_r + U_i = U$ that follows directly from the boundary conditions.

Secondly, if $R_S \to \infty$, $Y_S \to 0$ (our device is disconnected from the transmission line) $\bar{S} \to 0$, $I_0(t) \to 0$ and the coefficients $\bar{T}$ and $\bar{R}$ transform into the voltage transmission $\bar{T} \to T = 2Y_0/(Y_0 + Y_L)$ and reflection $\bar{R} \to R = (Y_0 - Y_L)/(Y_0 + Y_L)$ coefficients, respectively.

It is important to note that side by side with equation (20), equations (21) and (22) derived in the present paper constitute a self-consistent response of the superlattice to the incident waves. They allow one to find both emitted $U_L(t), I_L(t)$ and reflected $U_r(t), I_r(t)$ waveforms as functions of the incident monochromatic wave power $P_i$ and, consequently, completely define conversion efficiency of the terahertz harmonics generation in the superlattice.

We also make a point that in the limit of the vanishing dielectric relaxation time $\tau_d \to 0$ equations (13), (14) and (20) simplify to a system of two differential equations of first order describing the self-consistent interaction of a plane wave with the lateral superlattice [33, 34]. In this case the excitation of the damped Bloch oscillations appears to be a basic mechanism of the terahertz harmonics generation. In contrast, in the present paper we have a system of three first-order differential equations, with the dielectric relaxation term in the left-hand side of equation (20) having an inhibitory influence on the electron Bloch oscillations. This influence gives rise to a novel picture of the superlattice nonlinear ac response in the terahertz frequency band.

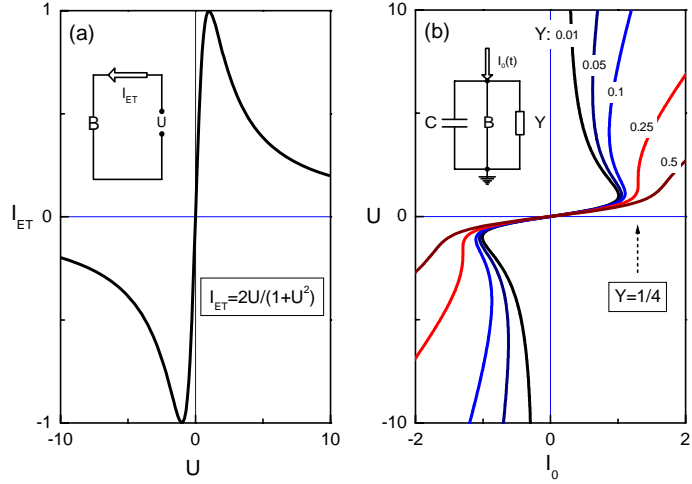

**Figure 2.** (a) The normalized Esaki-Tsu static current-voltage curve $I = I_{ET}(U)$. Inset: dc current flow when the dc voltage $U$ is directly applied to the superlattice two-terminal device. (b) The calculated effective voltage-current curve of the superlattice $U = U(I_0)$ for different values of the normalized radiation conductance $Y$. Inset: equivalent circuit for a superlattice device fed by the external ac current $I_0(t)$.

## 5. Relaxation oscillations

In a particular case of a monochromatic wave $V_i(t) = \bar{V}_i \cos(\omega t)$ incident on the superlattice the driving current $I_0(t)$ in equation (20) can be written as

$$I_0(t) = \bar{I}_0 \cos(\omega t) \qquad (28)$$

In terms of the incident power $P_i = (\bar{V}_i)^2 Y_0 / 2$ the amplitude of the current $\bar{I}_0$ can be expressed as

$$\bar{I}_0 = \bar{S}\left(\frac{P_i}{P_0}\right)^{1/2}, \qquad (29)$$

where the characteristic power $P_0 = i_p^2 / 8Y_0$ depends on the superlattice peak current $i_p$, the transmission line conductance $Y_0$, and the series resistance factor $\bar{S}$

Consider now that the frequency of the monochromatic wave is low enough, i.e. $\omega\tau \ll 1$, $\omega\tau_d \ll 1$. In this particular case equation (13) and (14) result in the normalized Esaki-Tsu static current-voltage curve $I_{ET} = 2U/(1+U^2)$ shown in figure 2(a). The inset in figure 2(a) illustrates schematically the dc current flow when the dc voltage $U$ is directly applied to the superlattice two-terminal device. In this case the dc current should follow the dc current –voltage curve which may be changed by effects of the space charge oscillations [35] caused by the negative differential conductivity effect ( $dI_{ET}/dU \leq 0$ ) in the region $U \geq 1$.

In our case described by equations (18)-(27) the current flow through the superlattice can be described by the equivalent circuit shown in the inset of figure 2(b). In terms of the normalized variables given above the equivalent circuit is presented by the electron current $B$ described by equations (13) and (14), the effective capacitance $C = \omega \tau_d$, and the normalized effective radiation conductance $Y$ connected in parallel with the electron current.

In the low-frequency limit the normalized superlattice voltage $U$ as a function of the driving current $I_0$ can be found from the equation

$$I_0 = I_{ET}(U) + YU . \tag{30}$$

Figure 2(b) shows the calculated $U = U(I_0)$ curves for different values of the normalized effective conductance $Y$. It is of critical importance that at

$$Y = (Y_0 + Y_L) R_0 S \leq \frac{1}{4} \tag{31}$$

the $U - I_0$ curve turns into the voltage-current curve of S-type in contrast to the N-type Esaki-Tsu dc current-voltage curve $I = I_{ET}(U)$. Equation (31) can be easily obtained analytically from equation (30) making use of the condition $dI_0/dU = 0$. In a qualitative sense the occurrence of the S-type $U - I_0$ curves can be easily explained if one accepts an idea that for the relatively small values of the superlattice dc resistance $R_0 \ll Z_0 Z_L / S(Z_0 + Z_L)$ the incident monochromatic wave mostly serves like an effective current (rather than a voltage) source.

The S-type effective current-voltage curve of the superlattice gives rise to excitation of the relaxation oscillations of the voltage at the load by the external driving current $I_0(t) = \overline{I}_0 \cos(\omega t)$. Really, at slow variation of the driving force $I_0(t)$ a sudden jump of the voltage $U$ should occur if $\overline{I}_0 \geq 1$ due to the multi-valued function $U(I_0)$. In terms of the incident power $P_i$ the criterion of the relaxation oscillations excitation $\overline{I}_0 = S(P_i/P_0)^{1/2} \geq 1$ can be written as

$$P_i \geq P_{th} = \frac{P_0}{S^2} . \tag{32}$$

In order to demonstrate the temporal evolution of the terahertz harmonics excitation we have performed a numerical simulation of the self-consistent response of our superlattice device making use of equations (13), (14) and (20). We take the circuit parameters as $Z_0 = 150$ Ω for the wave impedance of the ridged wave guide [36], and $Z_L = 377$ Ω as the wave impedance of free space.

For the numerical simulation we choose the superlattice parameters close to that ones described in experiments [37], i.e. the superlattice mesa diameter $D \approx 2.5$ μm, the peak current $i_p = 13$ mA, the peak current density $j_p \approx 270$ kA/cm$^2$, the superlattice period $d = 6.2$ nm, the miniband width $\Delta \approx 30$ meV, the density of electrons $n \approx 10^{18}$ cm$^{-3}$, the number of the superlattice periods $N \approx 20$, the peak electric field $E_p \approx 12$ kV/cm, the superlattice dc resistance $R_0 \approx 10$ $\Omega$, the series resistance $R_S \approx 10$ $\Omega$, the electron scattering frequency $\nu/2\pi \approx 1.9$ THz, the relative lattice dielectric constant $\varepsilon \approx 13$, and the dielectric relaxation frequency $f_d = 1/2\pi\tau_d \approx 3.4$ THz.

In this case for the effective radiation conductance $Y$, for the series resistance factor $S$, and for the characteristic power $P_0$ we obtain $Y \approx 0.085$, $S \approx 0.92$, and $P_0 \approx 3$ mW, respectively. Therefore, the criterion for occurrence of the S-type low-frequency voltage-current curve given by equation (31) is conservatively satisfied for samples described in [37]. We also note that the characteristic superlattice dc resistance $R_0 \approx 10$ $\Omega$ and the value of the input power $P_0 \approx 3$ mW are close to that ones used in the frequency multipliers described in papers [17-22].

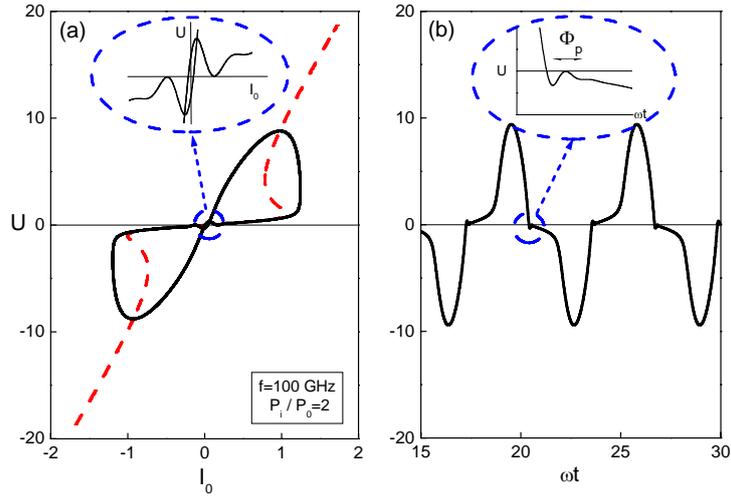

**Figure 3.** (a) Voltage-current phase plane demonstrating induced relaxation oscillations in a vertical superlattice device. (b) The calculated time sweep of the generated voltage $U$ as a function of the normalized time $\omega t$. Insets show the fragments of curves at low values of the generated voltage $U$.

Our simulation results are shown in figure 3 for the incident wave frequency $f = \omega/2\pi = 100$ GHz close to the frequencies chosen in experiments [17-22] where superlattice-based terahertz frequency multipliers have been explored. In this case the effective superlattice capacitance $C = \omega\tau_d$ and the effective relaxation time for electron scattering $\tau_s = \omega\tau$ are $C \approx 3\times 10^{-2}$ and $\tau_s \approx 5\times 10^{-2}$, respectively. The incident wave power $P_i$ is taken as $P_i/P_0 = 2$ corresponding to the condition of the relaxation oscillations excitation, i.e. to $\bar{I}_0 \approx 1.3$.

Figure 3(a) shows the $U - I_0$ phase plane demonstrating a stable limit cycle corresponding to the relaxation oscillations of the voltage $U$ exited in the superlattice by the external wave. The steady-state $U - I_0$ curve calculated from equation (30) is shown in this figure by the dashed line. Due to the inhibitory action of the superlattice effective capacitance $C$ instead of sudden jumps a smoothed variation of the voltage $U$ takes place. We have found that the limit cycle shape does not depend on initial conditions for $I$, $\eta$, and $U$. In the low-frequency range $\tau_s = \omega\tau <<1$ we also do not observe any kind of complicated electron dynamics such as a dissipative chaos [31] or a spontaneous dc current generation [32] in our system. We also make a point that no Bloch oscillations excitation is evident in figure 3 for the region of the superlattice parameters and frequency chosen for our numerical simulation.

The time sweep of the generated voltage $U$ as a function of the dimensionless time $\omega t$ is demonstrated in figure 3(b). As one can see from this figure, the waveform of the generated voltage $U$ is represented by a sequence of the short almost triangular pulses. In our simulation we found that the pulse front duration is strongly dependent on the value of the effective capacitance $C$. The pulse front becomes steeper with decreasing of the incident wave frequency $f$ and, consequently, with decreasing of the effective capacitance $C$.

It has been shown both theoretically and experimentally [35] that in n-doped superlattices the current oscillations at microwave frequencies can occur due to propagating dipole domains when the applied dc voltage $V_0$ exceeds the peak voltage $V_p$. The transient frequency of the propagating domains $f_{tr}$ was found to be $f_{tr} = 0.7 v_p / L$. In this context it is important to note that in our case the duration of the generated voltage pulses $T_{pulse}$ shown in figure 3(b) is small enough in comparison with the period of the incident wave, i.e. $\omega T_{pulse} \approx 1.2$, and $T_{pulse} \approx 2$ ps. This pulse duration is found to be much shorter than the transient time of the propagating domains $T_{tr} = 1/f_{tr} \approx 20$ ps for the typical values of the peak velocity $v_p \approx 10^6$ cm/s and for the superlattice length $L \approx 0.12$ μm [37] taken in our simulation. Consequently, we can conclude that the domain instability has no time to develop during the ultra short voltage pulse $U(t)$ applied across the superlattice.

## 6. Plasma frequency

The inset in figure 3(a) shows the fragment of the $U - I_0$ phase plane nearby the coordinate origin. One can see that the limit cycle demonstrates some specific oscillatory features in the region of small values of the generated voltage $U$. These features reveal themselves as strongly damped oscillations in the time sweep of the generated voltage shown in the inset of figure 3(b). We attribute this type of the temporal superlattice response to non-linear plasma oscillations induced by the external ac current $I_0(t)$. It is important to note that these oscillations form the pulse fronts and, consequently, define the generated harmonics power spectrum at high frequencies.

Actually, when a small amplitude electromagnetic wave is incident on the superlattice equations (13) and (14) result in the complex ac Drude conductivity. The Drude formula for the conductivity in combination with equation (20) yields:

$$\bar{U}_0 = \bar{I}_0 \sqrt{\frac{1+(\omega\tau)^2}{(2+Y-\omega^2\tau_D\tau)^2 + \omega^2(\tau_D+Y\tau)^2}}, \quad (33)$$

where $\bar{I}_0$ and $\bar{U}_0$ are the amplitudes of the sinusoidal current and voltage oscillations, respectively.

In the case of interest to us $Y \ll 1$ the amplitude of the voltage across the superlattice $\bar{U}_0$ demonstrates a resonance response at plasma frequency

$$\omega_p = \sqrt{\frac{2}{\tau_D \tau}} = \sqrt{\frac{2 j_p v}{\varepsilon\varepsilon_0 E_p}} = \sqrt{\frac{2ed j_p}{\varepsilon\varepsilon_0 \hbar}} \quad (34)$$

if the electron scattering time $\tau$ is large enough, i.e. if $\tau \gg \tau_D$.

The normalized period of the plasma oscillation $\Phi_p = \omega T_p$, where $T_p = 2\pi/\omega_p$, is shown in the inset of figure 3(b). The oscillation is strongly damped because the scattering frequency $v \approx 1.2 \times 10^{13} \text{s}^{-1}$ ($v/2\pi = 1.91$ THz) is close to the frequency of plasma oscillations $\omega_p \approx 2.3 \times 10^{13} \text{s}^{-1}$ ($\omega_p/2\pi = 3.7$ THz) for the case we simulated numerically.

We make a point that in our case the frequency of the incident wave $\omega$ is much smaller then the plasma frequency $\omega_p$, i.e. $\omega \ll \omega_p$. However, the plasma oscillations are excited due to a strongly non-linear response of our device to the external wave because a wide spectrum of harmonics can be internally generated in the superlattice due to the frequency multiplication effect. Therefore, the condition for the plasma oscillations excitation can be written as $k\omega = \omega_p$, where $k$ is the appropriate number of the generated harmonics.

## 7. Results and discussion

One can conclude from figure 3 that the generated voltage across the superlattice $U(t)$ is periodic in time, with the period $T$ being equal to the period of the incident wave $T = 2\pi/\omega$. In this case both the load voltage $U_L(\varphi)$, and the reflected voltage $U_r(\varphi)$ can be presented in terms of the Fourier expansion containing odd harmonics

$$U_{L,r}(\varphi) = \sum_k \left[ U_{L,r}^c(k)\cos(k\varphi) + U_{L,r}^s(k)\sin(k\varphi) \right], \quad (35)$$

where $k = 1, 3 ... \infty$, $U_{L,r}^c(k) = \pi^{-1} \int_{2\pi} U_{L,r}(\varphi) \cos(k\varphi) d\varphi$ and $U_{L,r}^s(k) = \pi^{-1} \int_{2\pi} U_{L,r}(\varphi) \sin(k\varphi) d\varphi$ are the coefficients of the Fourier expansion.

Therefore, for the emitted $P_L(k)$ and reflected $P_r(k)$ power spectrums one obtains:

$$P_L(k) = 4Y_0 Y_L R_0^2 \left[ U_L^c(k)^2 + U_L^s(k)^2 \right] P_0, \tag{36}$$

$$P_r(k) = 4Y_0^2 R_0^2 \left[ U_r^c(k)^2 + U_r^s(k)^2 \right] P_0. \tag{37}$$

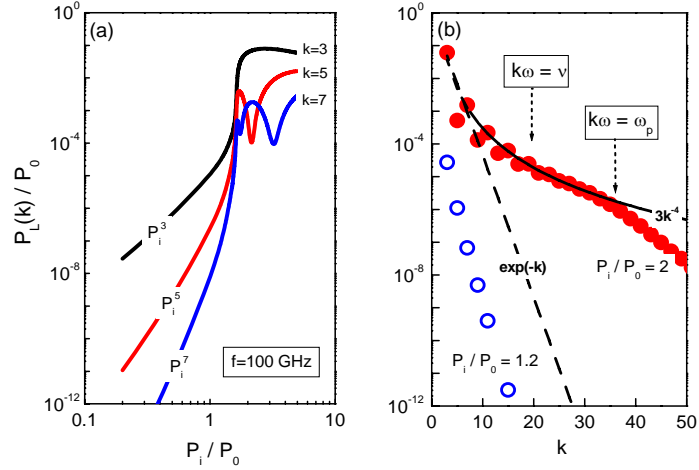

**Figure 4.** (a) The calculated normalized emitted harmonics power $P_L(k)/P_0$ as a function of the incident power $P_i/P_0$ for three different numbers of the emitted harmonics $k = 3, 5, 7$. (b) The calculated normalized emitted harmonics power $P_L(k)/P_0$ as a function of the harmonics number $k$ for the sub-threshold ($P_i/P_0 = 1.2$) and the super-threshold ($P_i/P_0 = 2$) values of the incident power $P_i$.

In figure 4(a) the emitted harmonics power $P_L(k)/P_0$ as a function of the incident power $P_i/P_0$ is shown for the third, fifth, and seventh harmonics of the incident wave. At low incident power $P_i/P_0 \ll 1$ we see that $P_L(k) \propto P_i^k$. This power law represents an evident consequence of the system's nonlinear response described in terms of a perturbation method. However, if the incident power $P_i$ exceeds the threshold value $P_i \geq P_{th}$ the emitted power $P_L(k)/P_0$ increases drastically. We attribute this effect to the relaxation oscillations excitation. At high values of the incident power $P_i/P_0 \geq 2$ the saturation and the oscillatory behavior of the emitted power $P_L(k)/P_0$ as a function of the incident power $P_i/P_0$ are seen in figure 4(a). From figure 4(a) we find that the maximum third harmonics conversion efficiency can be estimated as $\eta(3) = P_L(3)/P_i \approx 3\%$. The value of $\eta(3)$ is close to that one observed in papers [17, 18].

Figure 4(b) demonstrates the emitted harmonics power spectrum $P_L(k)/P_0$ as a function of the harmonics number $k$ for the sub-threshold and the super-threshold values of the incident power, i.e. for $P_i/P_0 = 1.2$ and $P_i/P_0 = 2$, respectively. Meanwhile at $P_i/P_0 = 1.2$ the emitted power $P_L(k)$ exponentially decays with increasing of the harmonics number $k$, at $P_i/P_0 = 2$ it obeys the $k^{-4}$ power law up to the roll-off frequency $\omega \approx \omega_r$. At this frequency the $k^{th}$ harmonics of the generated voltage $U(t)$ is close to the plasma frequency, i.e. $\omega_r = k\omega \approx \omega_p$. We also note the oscillatory behavior of the emitted power spectrum $P_L(k)/P_0$ at small harmonics numbers, i.e. at $k < \omega_p/\omega$.

The described features of the emitted harmonics power spectrum can be easily understood if one turns attention to the almost triangular shape of the generated pulses shown in figure 3(b). The analytical approximation of the generated pulses by a triangular shape and the calculation of the harmonics power spectrum according to equation (32) can qualitatively explain both the $k^{-4}$ decay of the harmonics power $P_L(k)$ as a function of $k$ and its oscillatory behavior at $k < \omega_p/\omega$.

For frequencies of the generated harmonics higher than the electron plasma frequency $k\omega > \omega_p$ a roll-off of the generated power $P_L(k)/P_0$ is clearly seen in figure 7(a). This behavior can be attributed to the time delay of the ac current response caused by relaxation terms presented in equations (13), (14) and (20). The presence of these relaxation terms gives rise to the damped plasma oscillations that finally form a front shape of the generated pulses and specify the roll-off frequency of the harmonics power spectrum $\omega_r = \omega_p$ belonging to the terahertz frequency band.

It is important to note that as it is seen from equation (34), the plasma frequency $\omega_p$ increases with increasing of the peak current density in the superlattice $j_p$ and, consequently, with increasing of the electron peak velocity $v_p$. Therefore, a superlattice electron device having higher miniband widths can produce broader spectrum of the generated harmonics.

We believe that the excitation of the relaxation oscillations described in the present paper may be relevant to experiments [17, 18], where the threshold-like dependence of the third harmonics power as a function of the incident power was observed. We would also like to mention that the ultra-broad spectrum of generated harmonics with the superlattice-based devices recently demonstrated in experiments [19-22] may also have a relevance to our calculations. We hope that the results obtained in the present article may contribute to a further development of the superlattice-based frequency multipliers for the high-resolution terahertz spectroscopy [19].

## 8. Conclusions

In conclusion, we have shown that generation of the ultra-short pulses by a vertical superlattice electron device can occur in a non-resonant circuit due to the relaxation oscillations excitation. In turn, this gives rise to the ultra-broad band spectrum of the emitted terahertz harmonics power. The relaxation oscillations excitation becomes feasible if the superlattice dc resistance is low enough in comparison with the characteristic radiation impedance of the external waveguide system. Our numerical simulations have shown that the dependence of the generated harmonics powers as a function of the incident wave power demonstrates a threshold-like behavior at a specific input power level. The Bloch oscillation excitations have been inherently involved in our simulation procedure making use of the balance equation approach describing electron transport in a superlattice. It has turned out that for typical superlattice parameters taken from experiments the roll-of frequency of the generated harmonics has no relevance to Bloch oscillations and is mostly specified by the plasma frequency of electrons in a superlattice miniband. We suggest that an increase of the superlattice miniband widths could essentially enhance both the efficiency and the spectral range of the generated terahertz harmonics.

**Acknowledgments**

We thank the Russian Foundation for Basic Research (RFBR) for support.

**References**

[1] Esaki L and Tsu R 1970 *IBM J. Res. Dev.* **14** 61-65
[2] Leo K 1998 *Semicond. Sci. Technol.* **13** 249-263
[3] Helm M 1995 *Semicond. Sci. Technol.* **10** 557-575
[4] Wacker A 2002 *Phys. Rep.* **357** 1-111
[5] Bonilla L L and Ghrahn 2005 *Rep. Prog. Phys.* **68** 577-683
[6] Willenberg H, Dohler G H and Faist J 2003 *Phys. Rev.* B **67** 085315-1-10
[7] Terazzi R, Gresch T, Giovanni M, Hoyler N, Sekine N and Faist J 2007 *Nat. Phys.* **3** 329-332
[8] Tonouchi M 2007 *Nat. Photon.* **1** 97-104
[9] Hyart T, Shorokhov A V and Alekseev K N 2007 *Phys. Rev. Lett.* **98** 220404-1-4
[10] Hyart T, Alekseev K N and Thuneberg E V 2008 *Phys. Rev.* B **77** 165330-1-13
[11] Hyart T, Alexeeva N V, Mattas J and Alekseev K N 2009 *Phys. Rev. Lett.* **102** 140405-1-4
[12] Ignatov A A and Romanov Yu A *1976 phys. stat. sol.* (b) **73** 327-333
[13] Lei X L 1997 *J. Appl. Phys.* **82** (2) 718-721
[14] Feise M W and Citrin D S 1999 *Appl. Phys. Lett.* **75** 3536-3538
[15] Schomburg E, Grenzer J, Hofbeck K, Dummer C, Winnerl S F, Ignatov A A, Renk K F, Pavel'ev D G, Koshurinov Yu I, Melzer B, Ivanov S, Ustinov V and Kop'ev P S 1996 *IEEE J. Sel. Top. Quantum Electron.* **2** 724-728
[16] Winnerl S, Schomburg E, Brandl S, Kus O, Renk K F, Wanke M C, Allen S J, Ignatov A A, Ustinov V, Zhukov A and Kop'ev P S 2000 *Appl. Phys. Lett.* **77** 1259-1261
[17] Klappenberger F, Renk K F, Rieder B, Pavelev D G, Ustinov V, Zhukov A, Maleev N and Vasilyev A 2004 *Appl. Phys. Lett.* **84** 3924-3926
[18] Renk K F, Stahl B I, Rogl A, Jansen T, Pavel'ev D G, Koshurinov Yu I, Ustinov V and Zhukov A 2005 *Phys. Rev. Lett.* **95** 126801-1-4
[19] Endres C P, Lewen F, Giesen T F, Schlemmer S, Paveliev D G, Koschurinov Y I, Ustinov V M, and Zhucov A E 2007 *Rev. Sci. Instr.* **78** 043106-1-6
[20] De Lucia F C 2010 *J. Mol. Spectrosc.* **261** 1-17
[21] Vaks V, Panin A, Pripolsin S and Paveliev D 2009 *Pros. International Workshop on Terahertz and mid Infrared Radiation: Basic Research and Practical Application* (3-6 November 2009 ITAP, Turunc-Marmaris, Turkey) p 89
[22] Khosropanah P, Baryshev A, Zhang W, Jellema W, Hovenier J N, Gao J R, T. M. Klapwijk T M, Paveliev D G, Williams B S, Kumar S, Hu Q, Reno J L, Klein B and Hesler J L *2009 Opt. Lett.* **34** 2958-2960
[23] Ignatov A A and Jauho A-P 1999 *J. Appl. Phys.* **85** 3643-3654
[24] Ghosh A W, Wanke M C, Allen S J and Wilkins J W 1999 *Appl. Phys. Lett.* **74** 2164-2166
[25] Brown E R and Parker C D 1996 *Phil. Trans. R. Soc. Lond.* A **354** 2365-2381

[26] Tretyakov M Yu, Volokhov S A, Golubyatnikov G Yu, Karyakin E N and Krupnov A F 1999 *Int. J. Infrared Mill. Waves* **20** 1443-1451

[27] Rizzi P A 1988 *Microwave Engineering. Passive Circuits* (New Jersey: Prentice-Hall) chapter 3

[28] Ignatov A A, Dodin E P and Shashkin V I 1991 *Mod. Phys. Lett.* B **5** 1087-1094

[29] Ignatov A A, Renk K F and Dodin E P 1993 *Phys. Rev. Lett.* **70** 1996-1999

[30] Brozak G, Helm M, DeRosa F, Perry C H, Kosa M, Bhat R and Allen Jr. S J 1990 *Phys. Rev. Lett.* **64** 3163-3166

[31] Alekseev K N, Berman G P, Campbell D K, Cannon E H and Cargo M C 1996 *Phys Rev.* B **54** 10625-10636

[32] Alekseev K N, Cannon E H, McKinney J C, Kusmartsev F V and Campbell D K 1998 *Phys. Rev. Lett.* **80** 2669-2672

[33] Dodin E P, Zharov A A and Ignatov A A 1998 *J. Exp. Theor. Phys.* **87** 1226-1234

[34] Dodin E P and Zharov A A 2003 *J. Exp. Theor. Phys.* **97** 127-134

[35] Schomburg E, Blomeier T, Hofbeck K, Grenzer J, Brandl S, Lingott I, Ignatov A A, Renk K F, pavel'ev D G, Koschurinov Yu, Melzer B Ya, Ustinov V M, Ivanov S V, Zhukov A and Kop'ev P S 1998 *Phys. Rev.* B **58** 4035-4038

[36] Mckay M and Helszajh J 1999 *IEEE Microwave Guided Wave Lett.* **9** 66-68

[37] Pavel'ev D G, Demarina N V, Koshurinov Yu I, Vasil'ev A P, Semenova E S, Zhukov A E and Ustinov V M 2004 *Semicond.* **38** 1105-1110